\documentclass{article}

\usepackage{PRIMEarxiv}

\usepackage[utf8]{inputenc}
\usepackage[T1]{fontenc}
\usepackage{hyperref}
\usepackage{url}
\usepackage{booktabs}
\usepackage{amsfonts}
\usepackage{nicefrac}
\usepackage{microtype}
\usepackage{fancyhdr}
\usepackage{graphicx}
\usepackage{mathtools}
\usepackage{amsmath}
\usepackage{subcaption}
\usepackage[table]{xcolor}
\usepackage{latexsym}
\usepackage{enumitem}

\newtheorem{theorem}{\bf Theorem}[section]
\newtheorem{lemma}{\bf Lemma}[section]

\newtheorem{definition}{\bf Definition}[section]
\newtheorem{conjecture}{\bf Conjecture}[section]
\newtheorem{assumption}{\bf Assumption}[section]

\numberwithin{equation}{section}
\def\qed{\hfill $\Box$}
\newcommand\fillb{\cellcolor{black!20}}
\mathtoolsset{showonlyrefs=false}
\hypersetup{
  colorlinks=true,
  linkcolor=blue,
  citecolor=blue
}
\usepackage[]{enumitem}
\setlist[enumerate]{
    labelsep=8pt,
    labelindent=0.5\parindent,
    itemindent=0pt,
    leftmargin=*,
    before=\setlength{\listparindent}{-\leftmargin},
}

\title{Optimal preference satisfaction for conflict-free joint decisions}

\author{
    Hiroaki Shinkawa$^{1, \dagger}$, Nicolas Chauvet$^1$, Guillaume Bachelier$^2$,\\
    André Röhm$^1$, Ryoichi Horisaki$^1$, and Makoto Naruse$^1$\\
    $^{1}$Graduate School of Information Science and Technology, The University of Tokyo, Tokyo, Japan\\
    $^{2}$ Université Grenoble Alpes, CNRS, Inst. Neel, Grenoble, France
}

\begin{document}
\maketitle

\begin{abstract}
    We all have preferences when multiple choices are available. If we insist on satisfying our preferences only, we may suffer a loss due to conflicts with other people's identical selections.
    Such a case applies when the choice cannot be divided into multiple pieces due to the intrinsic nature of the resources. Former studies, such as the top trading cycle, examined how to conduct fair joint decision-making while avoiding decision conflicts from the perspective of game theory when multiple players have their own deterministic preference profiles.
    However, in reality, probabilistic preferences can naturally appear in relation to the stochastic decision-making of humans.
    Here, we theoretically derive conflict-free joint decision-making that can satisfy the probabilistic preferences of all individual players.
    More specifically, we mathematically prove the conditions wherein the deviation of the resultant chance of obtaining each choice from the individual preference profile, which we call the loss, becomes zero, meaning that all players' satisfaction is perfectly appreciated while avoiding decision conflicts.
    Furthermore, even in situations where zero-loss conflict-free joint decision-making is unachievable, we show how to derive joint decision-making that accomplishes the theoretical minimum loss while ensuring conflict-free choices.
    Numerical demonstrations are also shown with several benchmarks.
\end{abstract}

\keywords{joint decision-making \and resource allocation \and game theory \and multi-armed bandit \and optimization}

\section{Introduction}
Since the industrial revolution, we have witnessed important technological advances, but we are now faced with the finiteness of resources, which leads to the necessity for strategies of competition or sharing for their allocation \cite{hira2017emergence, george2018management}.
We need to consider other people's, or any entities', choices along with those of ours. One of the critical issues we need to be concerned about is decision conflicts.
For example, suppose a lot of people or devices try to connect to the same mobile network simultaneously. In that case, the wireless resources available per person or device will be significantly smaller, resulting in poor communication bandwidth or even zero connectivity due to congestion \cite{liu2007resource}.
The example above considers a case where an individual suffers due to a choice conflict with others. Here, the division of resources among entities is allowed. However, in other cases, separable allocation to multiple entities is not permitted.
Think of a draft in a professional football league. Each of the clubs has specific players in mind to pick in the draft; however, in principle, only a single club can sign with one player. That is, decision conflicts must be strictly resolved.
These examples highlight the importance of accomplishing individual satisfaction while avoiding decision conflicts.

Indeed, Sharpley and Scarf proposed the Top Trading Cycle (hereinafter called TTC) allocation algorithm \cite{shapley1974cores} to address this problem.
A typical example is known as the house allocation problem. Each of the $n$ students ranks the houses they want to live in from a list of $n$ houses. In this situation, TTC allocates one house to each student.
The solution given by TTC is known to be game-theoretic core-stable; that is, arbitrary numbers of students can swap houses with each other and still not get a better allocation than the current one.
There are many other mechanisms which originate in TTC including those that allow indifference in preference profiles \cite{alcalde2011exchange, aziz2012housing, saban2013house} and those where agents can be allocated fractions of houses \cite{athanassoglou2011house}.

While TTC nicely works when players have deterministic rankings, humans or artificial machines can also have probabilistic preferences. That is, unlike in the house allocation problem, an agent with probabilistic preferences would be unsatisfied by always receiving its top preference.
Rather, over repeated allocations, the distribution of outcomes and their probabilistic preferences should become similar. Such probabilistic preferences can naturally appear in relation to the stochastic decision-making of humans.
The multi-armed bandit (MAB) problem \cite{robbins1952some, sutton2018reinforcement}, for example, typically encompasses probabilistic preferences.

In the MAB problem, one player repeatedly selects and draws one of several slot machines based on a certain policy. The reward for each slot machine is generated stochastically based on a probability distribution, which the player does not know a priori.
The purpose of the MAB problem is to maximize the cumulative reward obtained by the player under such circumstances.
First, the player draws all the machines, including those that did not generate many rewards, to accurately estimate the probability distribution of the reward that each machine follows. This aspect is called exploration.
On the other hand, after the player is confident in the estimation, the extensive drawing of the highest reward probability machine increases the cumulative reward.
This is called exploitation.
In order to successfully solve the MAB problem, it is necessary to strike a balance between exploration and exploitation especially in a dynamically changing environment, which is called the exploration-exploitation dilemma \cite{march1991exploration}.

We can see an example of probabilistic preferences in the context of the MAB problem. The softmax algorithm, one of the well-known stochastic decision-making methods, is an efficient way to solve this problem \cite{vermorel2005multi}. Specifically, if the empirical reward for each machine $i$ at a certain time $t$ is $\mu_i(t)$,
the probability that a player selects each machine $i$ is expressed as
\begin{equation}
    s_i(t) = \frac{e^{\mu_i(t)/\tau}}{\sum\limits_k e^{\mu_k(t)/\tau}}.
\end{equation}
Here, $\tau$ is called the \it{temperature}, \rm which controls the balance between exploration and exploitation. This means that the player has a probabilistic preference at each time step as to which machine to select and with what percent.

Furthermore, we can extend the MAB problem to the case involving multiple players, which is called the competitive multi-armed bandit problem (competitive MAB problem) \cite{lai2010cognitive, kim2016harnessing}.
In the competitive MAB problem, when multiple players simultaneously select the same machine and that machine generates a reward, the reward is distributed among the players.
The aim is to maximize collective rewards of the players while ensuring equality among them \cite{deneubourg1989collective}.
Decision conflicts inhibit the players from maximizing the total rewards in such situations.

Indeed, wireless communications, for example, suffer from such difficulties wherein the simultaneous usage of the same band by multiple devices results in performance degradation for individual devices.
In solving competitive MAB problems, Chauvet \it{et al.} \rm theoretically and experimentally demonstrated the usefulness of quantum entanglement to avoid decision conflicts without direct communication \cite{chauvet2019entangled, chauvet2020entangled}.
A powerful aspect of the usage of entanglement is that the optimization of the reward by one player leads to the optimization of the total rewards for the team. This is not easily achievable because normally, if everyone tries to draw the best machine, it leads to decision conflicts and thus diminishes the total rewards in the competitive MAB problem.
Moreover, Amakasu \it{et al.} \rm proposed to utilize quantum interference so that the number of choices can be extended to an arbitrary number while perfectly avoiding decision conflicts \cite{amakasu2021conflict}.
A more general example is multi-agent reinforcement learning \cite{tan1993multi, busoniu2008comprehensive}. In this case, multiple agents learn individually and make probabilistic choices while balancing exploration and exploitation at each time step.
Successfully integrating these agents and making cooperative decisions without selection conflicts can help accelerate the learning \cite{mnih2016asynchronous}.

With such motivations and backgrounds, what should be clarified is how the probabilistic preferences of individual players can be accommodated while avoiding choice conflicts.
More fundamentally, the question is whether it is really possible to satisfy all players' preferences while eliminating decision conflicts. If yes, what is the condition to realize such requirements?

This study theoretically clarifies the condition when joint decision-making probabilities exist so that all players' preference profiles are perfectly satisfied.
In other words, the condition that provides the loss, which is the deviation of each player's selection preference to the resulting chance of obtaining each choice determined via the joint decision probabilities, becomes zero is clearly formulated.
Furthermore, we derive the joint decision-making probabilities that minimize the loss even when such a condition is not satisfied.
In the present work, we purely focus on the satisfaction of the players' preferences, and the discussion which links to external environments such as rewards is left for future studies.

\section{Theory}\label{sec:construction method}
\subsection{Problem formulation}\label{subsec:problem}
This section introduces the formulation of the problem under study. To begin with, we will examine the cases where the number of players or agents is two, who are called player A and player B.
There are $N$ choices for each of them, which are called arms in the literature on the MAB problem, where $N$ is a natural number greater than or equal to 2.

\begin{figure}[t]
    \centering
    \includegraphics[width=12.0cm]{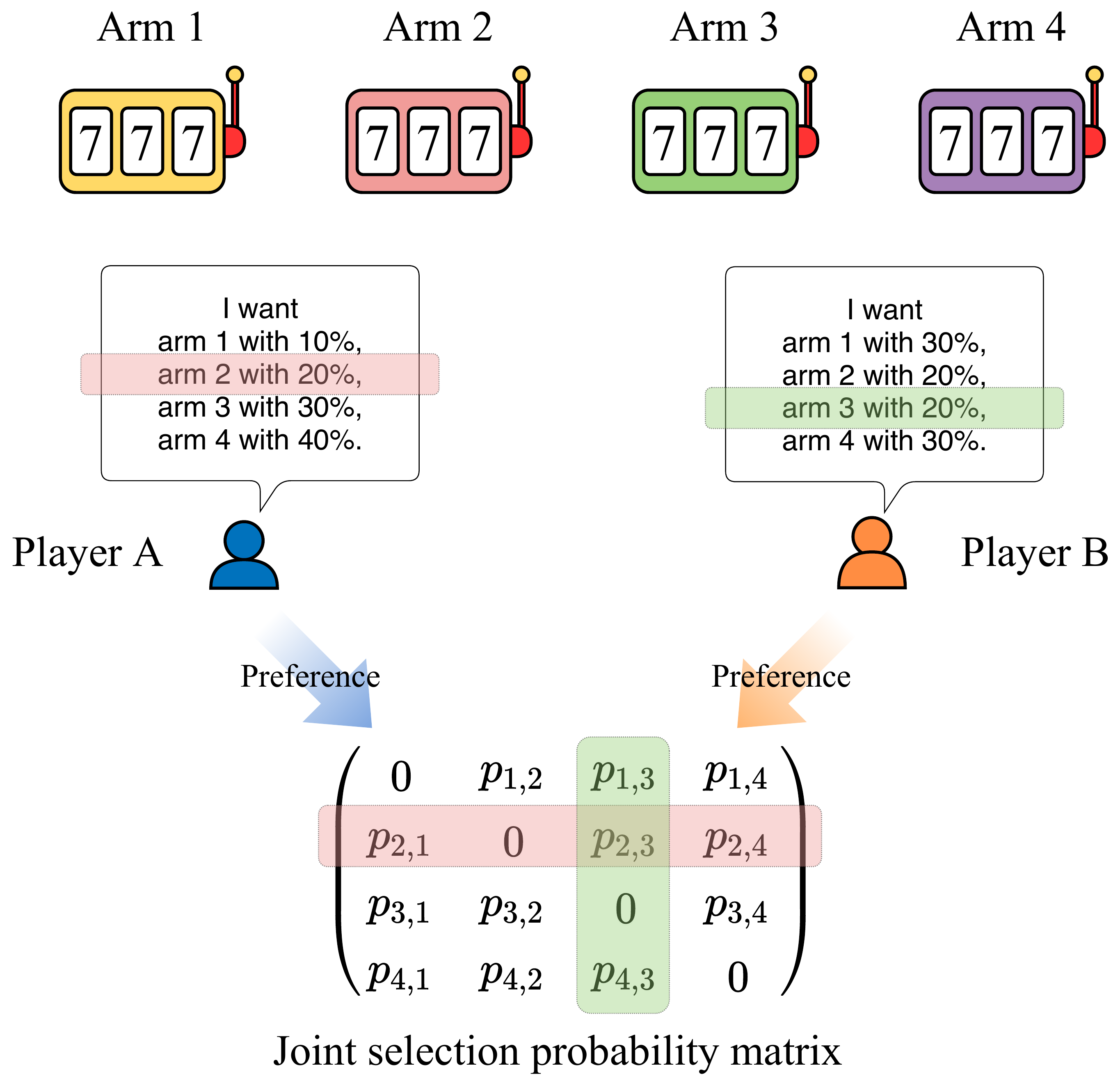}
    \caption{Problem settings. The sum of each row should be close to the corresponding preference of player A, and the sum of each column should be close to the corresponding preference of player B.}
    \label{fig:problem_settings}
\end{figure}

Each of players A and B selects a choice probabilistically depending on his or her preference probability.
Here, the preference of player A is represented by
\begin{equation}
    \boldsymbol{A} = \begin{pmatrix}A_1&A_2&\cdots&A_N\end{pmatrix}.
\end{equation}
Similarly, player B's preference is given by
\begin{equation}
    \boldsymbol{B} = \begin{pmatrix}B_1&B_2&\cdots&B_N\end{pmatrix}.
\end{equation}
These preferences have the typical properties of probabilities:
\begin{equation}
    A_1+A_2+\cdots+A_N = B_1+B_2+\cdots+B_N = 1, \quad A_i \geq 0, \quad B_i \geq 0\quad (i=1, 2, \cdots, N).
\end{equation}
The upper side of Figure \ref{fig:problem_settings} schematically illustrates such a situation where player A and B each has their own preference in a four-arm case.

Now, we introduce the joint selection probability matrix, which is given in the form of
\begin{equation}
    \boldsymbol{P} =
    \begin{pmatrix}
        0&p_{1,2}&\cdots&p_{1,N}\\
        p_{2,1}&0&\cdots&\vdots\\
        \vdots&\vdots&\ddots&\vdots\\
        p_{N,1}&\cdots&\cdots&0
    \end{pmatrix}.
\end{equation}
The non-diagonal element $p_{i,j}$, with $i\neq j$, denotes the probability of when player A selects arm $i$ and player B chooses arm $j$.
Since we consider non-conflict choices, the diagonal terms are all zero. Here, the summation of $p_{i,j}$ over all $i$s and $j$s is unity and $p_{i,j} \geq 0$.
\begin{equation}
    \sum_{i,j}p_{i,j} = 1, \quad p_{i,j} \geq 0.
\end{equation}

Our interest is which $\boldsymbol{P}$ meets the preferences of both players A and B. To incorporate such a perspective, we formulate the degree of satisfaction of the players in the following way.
By summing up $p_{i,j}$s row by row of the joint selection probability matrix, a preference profile given by
\begin{equation}
    \pi_A(i) = \sum_j p_{i,j}
\end{equation}
is obtained. $\pi_A(i)$ represents the probability of player A selecting arm $i$ as a result of the joint selection probability matrix $\boldsymbol{P}$.
We call $\pi_A(i)$ ``the satisfied preference'' (the link with the preference $A_i$ will be discussed hereafter).
\begin{equation}
    \sum_i \pi_A(i) = 1
\end{equation}
holds based on the definition of the joint selection probability matrix.
Such a structure is illustrated by the red shaded elements in the second row of the joint selection probability matrix in Figure \ref{fig:problem_settings}.

Similarly, the satisfied preference along the columns of $\boldsymbol{P}$
\begin{equation}
    \pi_B(j) = \sum_i p_{i,j}
\end{equation}
represents the probability of player B selecting arm $j$ as a result of the joint selection probability matrix $\boldsymbol{P}$.

Our aim is to find the optimal $p_{i,j}$s wherein
\begin{equation}
    \pi_A(i) \approx A_i,\quad \pi_B(j) \approx B_j
\end{equation}
hold for all $i=1, 2, \ldots, N$ and $j=1, 2, \ldots, N$; that is, the player's preferences $(A_i, B_j)$ is the same or close to the satisfied preferences $(\pi_A(i), \pi_B(j))$.

To quantify such a metric, we define the loss $L$ akin to an $L_2$-norm.
\begin{equation}
    L = \sum_i \left(\pi_A(i)-A_i\right)^2 + \sum_j \left(\pi_B(j)-B_j\right)^2,
\end{equation}
which comprises the sum of squares of the gap between the preferences and the satisfied preferences for the two players.
In the following sections, we prove that the loss defined above can be {\it{zero}}; in other words, the perfect satisfaction for the players can indeed be realized, under certain conditions. Furthermore, even in cases when zero-loss is not achievable, we can systematically derive the joint selection probability matrix that ensures the minimum loss.

Here, we define the popularity $S_i$ for each arm and propose three theorems and one conjecture based on the values of $S_i$.
\begin{definition}
    \normalfont The popularity $S_i$ is defined as the sum of the preferences of player A and player B for arm $i$.
    \begin{equation}
        S_i \coloneqq A_i + B_i \quad (i=1, 2, \ldots, N).
    \end{equation}
    Since the preferences $A_i$ and $B_i$ are probabilities, it holds that
    \begin{equation}
        \sum_i S_i = \sum_i A_i + \sum_i B_i = 2.
    \end{equation}
\end{definition}
Hereinafter, the minimum loss is denoted by $L_{\text{min}}$.
\begin{equation}
    L_{\text{min}} = \min_{\boldsymbol{P}}\{L\}.
\end{equation}
In the meantime, the loss function can be defined as other forms, such as the Kullback-Leibler (KL) divergence, also known as relative entropy \cite{kullback1951information}. The extended discussions on such metrics are future studies, but we consider that the fundamental structure of the problem under study will not change significantly.

\subsection{Theorem 1}\label{sbsec:theorem1}
\subsubsection{Statement}
\begin{theorem}\label{thm:0loss_case}
    Assume that all the popularities $S_i$ are smaller than or equal to 1. Then, it is possible to construct a joint selection probability matrix which makes the loss $L$ equal to zero.
    \begin{equation}
        \forall i; S_i \leq 1 \Rightarrow L_{\normalfont\text{min}} = 0.
    \end{equation}
\end{theorem}
\subsubsection{Auxiliary lemma}
To prove Theorem \ref{thm:0loss_case}, we first prove a lemma, where the problem settings are slightly modified.
In the original problem, we treat each player's preference as probabilities. Here, however, their preferences do not have to be probabilities as long as they are non-negative, and the sum of the preference for each player is given by a constant $T$, which we refer to as the total preference.
Their preferences are called $\hat{\boldsymbol{A}}$ and $\hat{\boldsymbol{B}}$ respectively. We will put a hat over each notation to avoid confusion about which definition we are referring to between the original problem and the modified problem.
\begin{gather}
    \hat{\boldsymbol{A}} = \begin{pmatrix}\hat{A}_1&\hat{A}_2&\cdots&\hat{A}_N\end{pmatrix}.\\
    \hat{\boldsymbol{B}} = \begin{pmatrix}\hat{B}_1&\hat{B}_2&\cdots&\hat{B}_N\end{pmatrix}.\\
    \hat{A}_1+\hat{A}_2+\cdots+\hat{A}_N = \hat{B}_1+\hat{B}_2+\cdots+\hat{B}_N = T, \quad \hat{A}_i \geq 0, \quad \hat{B}_i \geq 0 \quad (i=1, 2, \ldots, N). \label{definition:pref_hat}
\end{gather}
We still refer to
\begin{equation}
\hat{\boldsymbol{P}} =
    \begin{pmatrix}
        0&\hat{p}_{1,2}&\cdots&\hat{p}_{1,N}\\
        \hat{p}_{2,1}&0&\cdots&\vdots\\
        \vdots&\vdots&\ddots&\vdots\\
        \hat{p}_{N,1}&\cdots&\cdots&0
    \end{pmatrix}
\end{equation}
as a joint selection probability matrix even though, technically, each $\hat{p}_{i,j}$ is no longer a probability but a ratio of each joint selection happening, and the sum of all entries is $T$, instead of 1.
The other notations are defined similarly to the original problem.
\begin{gather}
    \text{The satisfied preference:}\qquad\hat{\pi}_A(i) = \sum_j \hat{p}_{i,j}, \quad \hat{\pi}_B(j) = \sum_i \hat{p}_{i,j},\\
    \text{The loss:}\qquad\hat{L} = \sum_i (\hat{\pi}_A(i)-\hat{A}_i)^2 + \sum_j (\hat{\pi}_B(j)-\hat{B}_j)^2, \quad \hat{L}_{\text{min}} = \min_{\hat{\boldsymbol{P}}}\{\hat{L}\},\\
    \text{The popularity:}\qquad\hat{S}_i=\hat{A}_i+\hat{B}_i.
\end{gather}
\begin{lemma}\label{lemma:0loss_case}
    Assume that all the popularities $\hat{S}_i$ are smaller than or equal to the total preference $T$. Then, it is possible to construct a joint selection probability matrix which makes the loss $\hat{L}$ equal to 0.
    \begin{equation}
        \forall i; \hat{S}_i \leq T \Rightarrow \hat{L}_{\normalfont\text{min}} = 0.
    \end{equation}
\end{lemma}
We see that Theorem \ref{thm:0loss_case} is a special case of Lemma \ref{lemma:0loss_case} when $T=1$.

\subsubsection{Outline of the proof}
We prove Lemma \ref{lemma:0loss_case} by mathematical induction. In sections \ref{sbsbsec:small_arms}, we show that Lemma \ref{lemma:0loss_case} holds for $N=2$ and $N=3$. In section \ref{sbsbsec:general_arms} and \ref{sbsbsec:construction}, we suppose that Lemma \ref{lemma:0loss_case} has been proven for $N$ arms, and then show that Lemma \ref{lemma:0loss_case} also holds for $N+1$ arms.
Specifically, in section \ref{sbsbsec:general_arms}, we assume the existence of $p_{i,j}$s which satisfy certain conditions and prove that the perfect satisfaction for the players is achievable using these $p_{i,j}$s. Then, in section \ref{sbsbsec:construction}, we verify their existence.

\subsubsection{When \texorpdfstring{$N=2$}{Lg} and when \texorpdfstring{$N=3$}{Lg}}\label{sbsbsec:small_arms}
When $N=2$,
\begin{equation}
    \hat{S}_1 = \hat{S}_2 = T
\end{equation}
is the only case where the assumption of Lemma \ref{lemma:0loss_case}, that is, $\forall i; \hat{S}_i \leq T$ is fulfilled. Due to the constraint \eqref{definition:pref_hat},
\begin{equation}
    \hat{A}_2 = T - \hat{A}_1, \quad \hat{B}_1 = T - \hat{A}_1, \quad \hat{B}_2 = \hat{A}_1.
\end{equation}
In this case,
\begin{equation}
    \hat{\boldsymbol{P}} = \begin{pmatrix}0&\hat{A}_1\\ \hat{A}_2 & 0\end{pmatrix}
\end{equation}
makes the loss equal to zero.

When $N=3$, by setting the values in the joint selection probability matrix as follows, the loss becomes zero.
\begin{equation}\label{opt_n_3}
    \hat{\boldsymbol{P}} = \begin{pmatrix}0&\hat{p}_{1,2}&-\hat{p}_{1,2}+\hat{A}_1\\T-\hat{p}_{1,2}-\hat{A}_3-\hat{B}_3&0&\hat{p}_{1,2}-\hat{A}_1+\hat{B}_3\\\hat{p}_{1,2}+\hat{A}_3-\hat{B}_2&-\hat{p}_{1,2}+\hat{B}_2&0\end{pmatrix}.
\end{equation}
The satisfied preferences for players A and B via $\hat{\boldsymbol{P}}$ result in $(\hat{A}_1, \hat{A}_2, \hat{A}_3)$ and $(\hat{B}_1, \hat{B}_2, \hat{B}_3)$, which perfectly match with the preferences of players A and B, respectively.

However, it should be noted that all the elements in $\hat{\boldsymbol{P}}$ must be non-negative.
\begin{equation}
    \begin{gathered}
        \hat{p}_{1,2} \geq 0, \quad -\hat{p}_{1,2}+\hat{A}_1 \geq 0, \quad T-\hat{p}_{1,2}-\hat{A}_3-\hat{B}_3 \geq 0,\\
        \hat{p}_{1,2}-\hat{A}_1+\hat{B}_3 \geq 0, \quad \hat{p}_{1,2}+\hat{A}_3-\hat{B}_2\geq 0, \quad -\hat{p}_{1,2}+\hat{B}_2 \geq 0.
    \end{gathered}
\end{equation}
To summarize these inequalities, the following inequality should hold;
\begin{equation}\label{main_ineq}
    \max{\{0,\hat{A}_1-\hat{B}_3,-\hat{A}_3+\hat{B}_2\}}\leq \hat{p}_{1,2} \leq \min{\{\hat{A}_1,\hat{B}_2, T-\hat{A}_3-\hat{B}_3\}}.
\end{equation}
If a non-negative $\hat{p}_{1,2}$ which satisfies \eqref{main_ineq} exists, $\hat{\boldsymbol{P}}$ given in \eqref{opt_n_3} is a valid joint selection probability matrix, which makes the loss zero.
In other words, if we can prove that $\hat{p}_{1,2}$ which is in the range of \eqref{main_ineq} exist, whatever $\hat{\boldsymbol{A}}$ and $\hat{\boldsymbol{B}}$ are, Lemma \ref{lemma:0loss_case} is proven to hold for $N=3$.

To prove that a non-negative $\hat{p}_{1,2}$ which satisfies \eqref{main_ineq} always exists, we consider the following three cases depending on the value of the left-hand side term of \eqref{main_ineq}.
In each case, we show that three candidates for the right-hand side term of \eqref{main_ineq} are greater than the left-hand side term.

\noindent\textbf{[Case 1]} When $\max{\{0,\hat{A}_1-\hat{B}_3,-\hat{A}_3+\hat{B}_2\}} = 0$.\\
$0 \leq \hat{A}_1, \quad 0 \leq \hat{B}_2$ are evident from the definition \eqref{definition:pref_hat}.
Also, $0 \leq T-\hat{A}_3-\hat{B}_3$ holds because of the assumption of the lemma $\hat{S}_3 \leq T$. Thus,
\begin{equation}\label{max_case1} 0 \leq \min{\{\hat{A}_1, \hat{B}_2, T-\hat{A}_3-\hat{B}_3\}}.\end{equation}

\noindent\textbf{[Case 2]} When $\max{\{0, \hat{A}_1-\hat{B}_3, -\hat{A}_3+\hat{B}_2\}} = \hat{A}_1-\hat{B}_3$.\\
First, the definition \eqref{definition:pref_hat} guarantees the following:
\begin{equation}\hat{A}_1-\hat{B}_3\leq \hat{A}_1.\end{equation}
Next, using the assumption $\hat{S}_1 \leq T$,
\begin{equation}\hat{A}_1-\hat{B}_3 \leq (T-\hat{B}_1)-\hat{B}_3 = B_2.\end{equation}
Finally,
\begin{equation}\hat{A}_1-\hat{B}_3 = (T-\hat{A}_2-\hat{A}_3)-\hat{B}_3 \leq T-\hat{A}_3-\hat{B}_3.\end{equation}
Therefore,
\begin{equation}\label{max_case2} \hat{A}_1-\hat{B}_3 \leq \min{\{\hat{A}_1,\hat{B}_2, T-\hat{A}_3-\hat{B}_3\}}.\end{equation}

\noindent\textbf{[Case 3]} When $\max{\{0, \hat{A}_1-\hat{B}_3, -\hat{A}_3+\hat{B}_2\}} = -\hat{A}_3+\hat{B}_2$.\\
$\hat{S}_2 \leq T$ implies the following.
\begin{equation}-\hat{A}_3+\hat{B}_2 \leq -\hat{A}_3+(T-\hat{A}_2) = A_1.\end{equation}
Also, the definition \eqref{definition:pref_hat} guarantees $-\hat{A}_3+\hat{B}_2 \leq \hat{B}_2$. Finally,
\begin{equation}-\hat{A}_3+\hat{B}_2 = -\hat{A}_3+(T-\hat{B}_1-\hat{B}_3) \leq T-\hat{A}_3-\hat{B}_3\end{equation}
holds because $\hat{B}_1 \geq 0$. Hence,
\begin{equation}\label{max_case3} -\hat{A}_3+\hat{B}_2 \leq \min{\{\hat{A}_1, \hat{B}_2, T-\hat{A}_3-\hat{B}_3\}}.\end{equation}

From \eqref{max_case1}, \eqref{max_case2} and \eqref{max_case3}, for any $\boldsymbol{A}, \boldsymbol{B}$,
\begin{equation}
    \max{\{0,\hat{A}_1-\hat{B}_3,-\hat{A}_3+\hat{B}_2\}}\leq \min{\{\hat{A}_1,\hat{B}_2, T-\hat{A}_3-\hat{B}_3\}}.
\end{equation}
Therefore, it is evident that $\hat{p}_{1,2}$ that satisfies \eqref{main_ineq} exist. Hence, $\hat{\boldsymbol{P}}$ given in \eqref{opt_n_3} is a valid joint selection probability matrix, which makes the loss $\hat{L}$ zero.

\subsubsection{Induction}\label{sbsbsec:general_arms}
In this section and the following section, we suppose that Lemma \ref{lemma:0loss_case} has been proven to hold when there are $N$ arms $(N \geq 3)$. Here, we show that Lemma \ref{lemma:0loss_case} also holds for $N+1$ case.
In the following argument, the arm with the lowest $\hat{S}_i$ is denoted by arm $K$.
\begin{equation}
    \min{\{\hat{S}_i\}} = \hat{S}_K.
\end{equation}
Now, we make the following assumption, which focuses on only the values in the $K$th row or the $K$th column.
\begin{assumption}\label{assumption:Kth_arm}
    There exist $\hat{p}_{K,1}, \hat{p}_{K,2}, \ldots, \hat{p}_{K,N+1}, \hat{p}_{1,K}, \hat{p}_{2,K}, \ldots, \hat{p}_{N+1,K}$ which satisfy all the following conditions \eqref{eq_a}--\eqref{ineq_ab}.\\

    The sum of the $K$th row is equal to $\hat{A}_K$.
    \begin{equation}\label{eq_a}\sum_{j=1}^{N+1} \hat{p}_{K,j} = \hat{A}_K.\end{equation}

    The sum of the $K$th column is equal to $\hat{B}_K$.
    \begin{equation}\label{eq_b}\sum_{i=1}^{N+1} \hat{p}_{i,K} = \hat{B}_K.\end{equation}

    The sum of the $j$th column without the $K$th row is non-negative.
    \begin{equation}\label{ineq_a}\sum_{i \neq K} \hat{p}_{i,j} \geq 0 \Leftrightarrow \hat{p}_{K,j} \leq \hat{B}_j \quad (j=1, 2, \ldots, K-1, K+1, \ldots, N+1).\end{equation}

    The sum of the $i$th row without the $K$th column is non-negative.
    \begin{equation}\label{ineq_b}\sum_{j \neq K} \hat{p}_{i,j} \geq 0 \Leftrightarrow \hat{p}_{i,K} \leq \hat{A}_i \quad (i=1, 2, \ldots, K-1, K+1, \ldots, N+1).\end{equation}

    In the gray-shaded area of the joint selection probability matrix below, all the remaining popularities are smaller than or equal to the remaining total preference. Note that $\hat{S}_K = \min\{\hat{S}_i\}$.
    \begin{equation}\label{ineq_ab}(\hat{A}_i-\hat{p}_{i,K})+(\hat{B}_i-\hat{p}_{K,i}) \leq T-\hat{S}_K \quad (i=1, 2, \ldots, K-1, K+1, \ldots, N+1).\end{equation}
\end{assumption}
We first suppose that Assumption \ref{assumption:Kth_arm} is valid and show the way to construct a joint selection probability matrix which makes the loss $\hat{L}$ zero, using these $\hat{p}_{i,j}$s. Later, we will prove that Assumption \ref{assumption:Kth_arm} is indeed correct.

Here, we show that, using the $\hat{p}_{i,j}$s which are assumed to exist in Assumption \ref{assumption:Kth_arm}, we can make the loss $\hat{L}$ equal to zero.
From conditions \eqref{eq_a} and \eqref{eq_b}, $(\pi_A(K)-A_K)^2+(\pi_B(K)-B_K)^2$, which are terms in the definition of $\hat{L}$ regarding arm $K$, is zero.
\begin{equation}\label{loss_Kth_arm}
    (\pi_A(K)-A_K)^2+(\pi_B(K)-B_K)^2 = 0.
\end{equation}
Next, we consider the loss for the remaining part of the joint selection probability matrix in the gray-shaded region described in \eqref{remaining_part}, which is defined by
\begin{equation}
    \hat{L}_\text{rem} = \hat{L} - \{(\hat{\pi}_A(K)-\hat{A}_K)^2+(\hat{\pi}_B(K)-\hat{B}_K)^2\}.
\end{equation}
\begin{equation}\label{remaining_part}
    \left(\begin{array}{ccccccc}
        \fillb 0&\fillb *&\fillb\cdots&p_{1,K}&\fillb\cdots&\fillb *&\fillb *\\
        \fillb *&\fillb 0&\fillb\cdots&p_{2,K}&\fillb\cdots&\fillb *&\fillb *\\
        \fillb\vdots&\fillb\vdots&\fillb\ddots&\vdots&\fillb\cdots&\fillb\vdots&\fillb\vdots\\
        p_{K,1}&p_{K,2}&\cdots&0&\cdots&p_{K,N}&p_{K,N+1}\\
        \fillb\vdots&\fillb\vdots&\fillb\vdots&\vdots&\fillb\ddots&\fillb\vdots&\fillb\vdots\\
        \fillb *&\fillb *&\fillb\cdots&p_{N,K}&\fillb\cdots&\fillb 0&\fillb *\\
        \fillb *&\fillb *&\fillb\cdots&p_{N+1,K}&\fillb\cdots&\fillb *&\fillb 0\\
    \end{array}\right)
\end{equation}
\begin{table}[t]
    \centering
    \caption{The preference setting which the remaining part of the joint selection probability matrix in the gray-shaded region follows. $\hat{A}_i^* = \hat{A}_i - \hat{p}_{i,K}, \quad \hat{B}_j^* = \hat{B}_j - \hat{p}_{K,j}$.}\label{tb:problem_settings_for_proof}
    \scalebox{0.9}{
        \begin{tabular}{|c||c|c|c|c|c|c|c||c|}
            \hline
            Player&Arm 1&Arm 2&$\cdots$&Arm $(K-1)$&Arm $(K+1)$&$\cdots$&Arm $(N+1)$&Sum\\
            \hline
            A&$\hat{A}_1^*$&$\hat{A}_2^*$&$\cdots$&$\hat{A}_{K-1}^*$&$\hat{A}_{K+1}^*$&$\cdots$&$\hat{A}_{N+1}^*$&$T-\hat{S}_K$\\
            B&$\hat{B}_1^*$&$\hat{B}_2^*$&$\cdots$&$\hat{B}_{K-1}^*$&$\hat{B}_{K+1}^*$&$\cdots$&$\hat{B}_{N+1}^*$&$T-\hat{S}_K$\\
            \hline
        \end{tabular}
    }
\end{table}
$\hat{L}_\text{rem}$ is the loss which corresponds to a problem where the preference setting is described in Table \ref{tb:problem_settings_for_proof}. $\hat{A}_i^*$ and $\hat{B}_i^*$ are defined by
\begin{equation}\hat{A}_i^* = \hat{A}_i - \hat{p}_{i,K}, \quad \hat{B}_j^* = \hat{B}_j - \hat{p}_{K,j}.\end{equation}
Now, we prove that the optimal $\hat{L}_\text{rem}$ is zero.
$A_i^*$ and $B_i^*$ fulfill the requirements for the preference; that is, the preferences need to be non-negative, because $p_{i,K}$s and $p_{K,j}$s follow \eqref{ineq_a} and \eqref{ineq_b} and the total preferences for both players are the same.
In this section, we suppose we have proven that Lemma \ref{lemma:0loss_case} holds when there are $N$ arms,
and because of \eqref{ineq_ab}, all the remaining popularities $\hat{S}_i^* \coloneqq \hat{A}_i^*+\hat{B}_i^* = (\hat{A}_i-\hat{p}_{i,K})+(\hat{B}_i-\hat{p}_{K,i})$ are smaller than or equal to the remaining total preference $T-\hat{S}_K$.
Thus, Lemma \ref{lemma:0loss_case} in the case of $N$ arms verifies
\begin{equation}\label{loss_other_arms}
    \min\{\hat{L}_\text{rem}\} = 0.
\end{equation}
From \eqref{loss_Kth_arm} and \eqref{loss_other_arms}, using $\hat{p}_{K,1}, \hat{p}_{K,2}, \ldots, \hat{p}_{K,N+1}, \hat{p}_{1,K}, \hat{p}_{2,K}, \ldots, \hat{p}_{N+1,K}$ and the optimal $\hat{p}_{i,j}$s for the gray-shaded region,
\begin{equation}
    \hat{L}_{\text{min}} = \min\{\hat{L}_\text{rem}\} + \{(\hat{\pi}_A(K)-\hat{A}_K)^2+(\hat{\pi}_B(K)-\hat{B}_K)^2\} = 0.
\end{equation}
Therefore, if Assumption \ref{assumption:Kth_arm} is correct, Lemma \ref{lemma:0loss_case} is proven to hold for $N+1$ arms.

\subsubsection{Verification of Assumption \ref{assumption:Kth_arm}}\label{sbsbsec:construction}
In this section, we prove that Assumption \ref{assumption:Kth_arm} is indeed correct.
Let us call the arm with the highest $\hat{S}_i$ arm $V$ ($V \neq K$).
\begin{equation}
    \max \{\hat{S}_i\} = \hat{S}_V
\end{equation}
First, we prove by contradiction that there is at most one arm which violates
\begin{equation}\label{already_ineq_ab}
    \hat{S}_i \leq T - \hat{S}_K,
\end{equation}
and if there is such an arm, arm $V$ is the one that breaks \eqref{already_ineq_ab}. Note that all the other arms, which satisfy \eqref{already_ineq_ab}, follow \eqref{ineq_ab}.
We assume that there are two arms which do not satisfy \eqref{already_ineq_ab} (we call them arm $V_1$ and arm $V_2$) and we will prove that this assumption leads to a contradiction.
\begin{equation}
    \hat{S}_{V_1} > T - \hat{S}_K, \quad \hat{S}_{V_2} > T - S_K.
\end{equation}
If we add each side,
\begin{equation}\label{ineq_s}
\hat{S}_{V_1} + \hat{S}_{V_2} > 2T - 2\hat{S}_K \Leftrightarrow 2T < \hat{S}_{V_1}+\hat{S}_{V_2}+2\hat{S}_K.
\end{equation}
Since $\min{\{\hat{S}_i\}}=\hat{S}_K$ and $N\geq 3$,
\begin{equation}
    \hat{S}_{V_1}+\hat{S}_{V_2}+2\hat{S}_K \leq \sum_{i=1}^{N+1} \hat{S}_i = 2T.
\end{equation}
Together with \eqref{ineq_s}, we obtain a contradiction
\begin{equation}
    2T < 2T.
\end{equation}
Therefore, it is proven, by contradiction, that there is at most one arm which violates \eqref{already_ineq_ab}.
Such an arm follows
\begin{equation}
    \hat{S}_i > T - \hat{S}_K,
\end{equation}
thus it is the one that has the highest $\hat{S}_{i}$, which we call arm $V$.

Now, we can systematically determine $\hat{p}_{i,j}$s in the $K$th row or the $K$th column to show Assumption \ref{assumption:Kth_arm} is correct; that is, these $\hat{p}_{i,j}$s fulfill the conditions \eqref{eq_a}--\eqref{ineq_ab}.
We cannot arbitrarily fill in the values in the $K$th row or the $K$th column so that their sum is $\hat{A}_K$ and $\hat{B}_K$, respectively, but have to take the other arms' preferences into consideration.
\eqref{ineq_a}, \eqref{ineq_b} and \eqref{ineq_ab} give us boundaries for $\hat{p}_{K,j}$s and $\hat{p}_{i,K}$s to exist.
Not only must we ensure that $\hat{p}_{K,j}$ and $\hat{p}_{i,K}$ do not exceed these boundaries, but we must also ensure that sufficient probability is assigned to the most popular arm $V$, because otherwise the left-hand side of \eqref{ineq_ab} can sometimes be greater than the right-hand side for arm $V$.
We consider the following three cases depending on size relation of $\hat{A}_K, \hat{B}_V, \hat{B}_K, \hat{A}_V$, and for each case, it is possible to construct the optimal $\hat{p}_{i,j}s$ that satisfy all the conditions \eqref{eq_a}--\eqref{ineq_ab}.
Note that these three cases are collectively exhaustive.
\begin{enumerate}[label=\textbf{[Case \arabic*]}]
    \item $\hat{A}_K \leq \hat{B}_V$ and $\hat{B}_K \leq \hat{A}_V$
    \item $\hat{A}_K > \hat{B}_V$
    \item $\hat{B}_K > \hat{A}_V$
\end{enumerate}
As a reminder, arm $K$ is the arm with the lowest popularity and arm $V$ is the one with the highest popularity.
For detailed construction procedures, see Appendix A.

With the above three cases, it is proven that $\hat{p}_{K,1}, \hat{p}_{K,2}, \ldots, \hat{p}_{K,N+1}, \hat{p}_{1,K}, \hat{p}_{2,K}, \ldots, \hat{p}_{N+1,K}$ which satisfy \eqref{eq_a}--\eqref{ineq_ab} always exist. In other words, Assumption \ref{assumption:Kth_arm} is indeed correct.
Therefore, if we assume that Lemma \ref{lemma:0loss_case} has been proven to hold for $N$ arms, then we know that Lemma \ref{lemma:0loss_case} holds when there are $N+1$ arms.

Hence, with mathematical induction, Lemma \ref{lemma:0loss_case} holds for any number of arms.
\qed

\subsection{Theorem 2}\label{sbsec:theorem2}
\subsubsection{Statement}
Here, we come back to the original problem where the players' preferences are probabilities. Note that the popularity $S_i$ still can be greater than 1 because it is the sum of the preferences for each arm.
\begin{theorem}\label{thm:non0loss_case}
    If any value of $S_i$ is greater than 1, it is not possible to make the loss $L$ equal to zero.\\
    In a case when the $N$th arm is the most popular; that is, $\max{\{S_i\}}=S_N>1$, the minimum loss is
    \begin{equation}
        L_{\text{min}} = \frac{N}{2(N-1)}\cdot (S_N-1)^2.
    \end{equation}
    The following joint selection probability matrix is one of the matrices which minimize the loss.
    \begin{equation}
        \tilde{\boldsymbol{P}} =
            \begin{pmatrix}
                0&0&\cdots&0&A_1+\epsilon\\
                0&0&\cdots&0&A_2+\epsilon\\
                \vdots&\vdots&\ddots&\vdots&\vdots\\
                0&0&\cdots&0&A_{N-1}+\epsilon\\
                B_1+\epsilon&B_2+\epsilon&\cdots&B_{N-1}+\epsilon&0
            \end{pmatrix}, \quad \epsilon=\frac{S_N-1}{2(N-1)}.
    \end{equation}
\end{theorem}
\subsubsection{Outline of the proof}
We can prove that the loss function $L$ is convex by showing that its Hessian is semidefinite. Details of the proof can be found in Appendix B.
In section \ref{sbsbsec:KKT}, we rewrite the problem into an optimization problem with an equality constraint and inequality constraints, and derive $\tilde{p}_{i,j}$s which satisfy the Karush-Kuhn-Tucker conditions \cite{karush1939minima, kuhn1951nonlinear} (hereinafter called KKT conditions). In a convex optimization problem, where the objective function and all the constraints are convex,
points which satisfy the KKT conditions give us the global optima \cite{boyd2004convex}. Therefore, the optimal joint selection probability matrix consists of the above-mentioned $\tilde{p}_{i,j}$s.

\subsubsection{KKT conditions}\label{sbsbsec:KKT}
Here, we derive the optimal joint selection probability matrix and calculate the minimum loss. The flattened vector of a joint selection probability matrix is defined as
\begin{equation}
    \boldsymbol{p} = \begin{pmatrix}p_{1,2}&p_{1,3}&\cdots&p_{1,N}&p_{2,1}&p_{2,3}&\cdots&p_{2,N}&\cdots&p_{N,N-1}\end{pmatrix}^T
\end{equation}
If functions $h$ and $g_{i,j}$s are defined as follows:
\begin{equation}
    h(\boldsymbol{p}) = \sum_{i,j} p_{i,j}-1, \quad g_{i,j}(\boldsymbol{p}) = -p_{i,j}\quad (i\neq j),
\end{equation}
the problem of minimizing the loss while satisfying the constraints can be written as
\begin{equation}
    \begin{aligned}
    \min_{\boldsymbol{p}} \quad & L(\boldsymbol{{p}})\\
    \textrm{s.t.} \quad & h(\boldsymbol{p}) = 0, \quad g_{i,j}(\boldsymbol{p})\leq 0.
    \end{aligned}
\end{equation}
Here, $L(\boldsymbol{p})$ is the loss which corresponds to $\boldsymbol{p}$. Since the objective function and the constraints are all convex, $\tilde{\boldsymbol{p}}$ which satisfies the KKT conditions below gives the global minimum.
\begin{equation}\label{KKT}
    \begin{gathered}
        \nabla L(\tilde{\boldsymbol{p}}) + \sum_{i,j} \lambda_{i,j} \nabla g_{i,j}(\tilde{\boldsymbol{p}}) + \mu \nabla h(\tilde{\boldsymbol{p}}) = \boldsymbol{0},\\
        \lambda_{i,j} g_{i,j}(\tilde{\boldsymbol{p}}) = 0,\\
        \lambda_{i,j} \geq 0,\\
        g_{i,j}(\tilde{\boldsymbol{p}}) \leq 0, \quad h(\tilde{\boldsymbol{p}})=0.
    \end{gathered}
\end{equation}
The following parameters satisfy all the conditions described in \eqref{KKT}.
\begin{equation}
    \begin{gathered}
        \epsilon = \frac{S_N-1}{2(N-1)},\\
        \mu = 2(N-2)\epsilon ,\\
        \lambda_{i,j} = \left\{
                \begin{array}{cl}
                2N\epsilon & (i\neq N \land j\neq N) \\
                0 & (otherwise)
                \end{array}
        \right.,\\
        \tilde{p}_{i,j} = \left\{
            \begin{array}{cl}
            0 & (i\neq N \land j\neq N) \\
            A_i + \epsilon & (j=N)\\
            B_j + \epsilon & (i=N)
            \end{array}
        \right.
    \end{gathered}
\end{equation}
See Appendix C for the verification of each condition.

Hence,
\begin{equation}
    \tilde{\boldsymbol{P}} =
    \begin{pmatrix}
    0&0&\cdots&0&A_1+\epsilon\\
    0&0&\cdots&0&A_2+\epsilon\\
    \vdots&\vdots&\ddots&\vdots&\vdots\\
    0&0&\cdots&0&A_{N-1}+\epsilon\\
    B_1+\epsilon&B_2+\epsilon&\cdots&B_{N-1}+\epsilon&0
    \end{pmatrix}, \quad \epsilon=\frac{S_N-1}{2(N-1)}
\end{equation}
gives the global optima. The minimum loss is
\begin{align}
    \begin{split}
        L_{\text{min}} &= \sum_i (\pi_A(i)-A_i)^2 + \sum_j (\pi_B(j)-B_j)^2\\
        &= \sum_{i=1}^{N-1}(\pi_A(i)-A_i)^2 + \sum_{j=1}^{N-1}(\pi_B(j)-B_j)^2 + (\pi_A(N)-A_N)^2 + (\pi_B(j)-B_N)\\
        &= \sum_{i=1}^{N-1}\epsilon^2 + \sum_{j=1}^{N-1}\epsilon^2 + \{-(N-1)\epsilon\}^2 + \{-(N-1)\epsilon\}^2\\
        &= (N-1)\epsilon^2 + (N-1)\epsilon^2 + (N-1)^2\epsilon^2 + (N-1)^2\epsilon^2\\
        &= \frac{N}{2(N-1)}\cdot (S_N-1)^2.
    \end{split}
\end{align}
Note that $\tilde{\boldsymbol{P}}$ is one example of the global optima and there could be other matrices which give us the same minimum loss.

Therefore, Theorem \ref{thm:non0loss_case} is proven.
\qed

\subsection{Theorem 3}
So far, we have presented two theorems concerning two players and $N$ arms. In this section, we consider the case where $M$ players exist where $M$ is greater than two.
The players are called player A, player B, player C and the like. Moreover, the preference of player $X$ for arm $i$ is denoted as $X^\dagger_i$. Similar to the original problem,
\begin{equation}
    \sum_i X^\dagger_i = 1, \quad X^\dagger_i \geq 0
\end{equation}
holds.
The sum of the preferences for each arm is denoted by the popularity
\begin{equation}
    S^\dagger_i = A^\dagger_i + B^\dagger_i + C^\dagger_i + \cdots.
\end{equation}
We define $d_x$ as the arm index which the $x$th player selects. Then, $p^\dagger_{d_1, d_2, \ldots, d_M}$ represents the joint selection probability of the $x$th player selecting arm $d_x$.
The collection of the joint selection probabilities is called the joint selection probability tensor.
The satisfied preference or the resultant selection probability for player $X$ consists of the sum of all the probabilities of cases where the $x$th player selects arm $i$:
\begin{equation}
    \pi^\dagger_x(i) = \underbrace{\sum_{\substack{d_1 \notin \{i\}}}\sum_{\substack{d_2 \notin \{i, d_1\}}}\cdots \sum_{\substack{d_M \notin \{i, d_1, d_2, \ldots , d_{M-1}\}}}}_{\text{Summation without the } x \text{th player}} p^\dagger_{d_1, d_2, \ldots, d_x=i , \ldots , d_M}.
\end{equation}
The loss is $L^\dagger$ is defined as the sum of squares of the gap between the preferences and the satisfied preferences.
\begin{equation}
    L^\dagger = \sum_x \sum_i \left(\pi^\dagger_x(i)-X^\dagger_i\right)^2.
\end{equation}
Here, the $x$th player is called player $X$.

\subsubsection{Statement}
\begin{theorem}\label{thm:general_non0loss}
    Suppose there are $M$ players and $N$ arms.
    If any of $S^\dagger_i$ is greater than 1, it is not possible to make the loss $L^\dagger$ equal to zero.
\end{theorem}

\subsubsection{Proof by contradiction}
Suppose, without loss of generality, that $S^\dagger_1 > 1$.
We first assume it is possible to make the loss $L^\dagger$ equal to zero, and then we will prove that it leads to a contradiction. If the loss $L^\dagger$ were to become zero, the followings are required.
\begin{equation}\label{sumeq_a}
    A^\dagger_1 = \sum_{\substack{d_2\notin \{1\}}}\sum_{\substack{d_3\notin \{1, d_2\}}} \cdots \sum_{\substack{d_M\notin \{1, d_2, d_3, \ldots , d_{M-1}\}}}p^\dagger_{1, d_2, d_3, \ldots , d_M}.
\end{equation}
\begin{equation}\label{sumeq_b}
    B^\dagger_1 = \sum_{\substack{d_1\notin \{1\}}}\sum_{\substack{d_3\notin \{d_1,1\}}} \cdots \sum_{\substack{d_M\notin \{d_1, 1, d_3, \ldots , d_{M-1}\}}} p^\dagger_{d_1, 1, d_3, \ldots , d_M}.
\end{equation}
\begin{equation}\label{sumeq_c}
    C^\dagger_1 = \sum_{\substack{d_1\notin \{1\}}}\sum_{\substack{d_2\notin \{d_1,1\}}} \cdots \sum_{\substack{d_M\notin \{d_1, d_2, 1, \ldots , d_{M-1}\}}} p^\dagger_{d_1, 1, d_3, \ldots , d_M}.
\end{equation}
\begin{equation*}
    \vdots
\end{equation*}
The terms which appear in the right-hand side of \eqref{sumeq_a} is not identical to the terms in the right-hand side of \eqref{sumeq_b} because $d_1 \neq 1$ in \eqref{sumeq_b}. Similarly, any term which appears in the right-hand side of \eqref{sumeq_a} is not identical to a variable which appear in the right-hand side of
all the equations \eqref{sumeq_b}, \eqref{sumeq_c}, $\ldots$. Therefore, if we take the sum of the right-hand sides of \eqref{sumeq_a}, \eqref{sumeq_b}, \eqref{sumeq_c}, $\ldots$, it should be smaller than or equal to the sum of all the elements in the joint selection probability tensor, which is 1.

On the other hand, if we take the sum of the left-hand sides of \eqref{sumeq_a}, \eqref{sumeq_b}, \eqref{sumeq_c}, $\cdots$,
we get $A^\dagger_1+B^\dagger_1+C^\dagger_1+\cdots = S^\dagger_1$. Thus, if we compare the both sides, we get $S^\dagger_1\leq 1$. This consequence contradicts with the assumption that $S^\dagger_1 > 1$.

Hence, with proof by contradiction, Theorem \ref{thm:general_non0loss} holds for any number of players and arms.
\qed

\subsection{Conjecture}
The construction method introduced in Theorem \ref{thm:0loss_case} is expected to be applicable to general $M$ players. Here, we propose a conjecture:
\begin{conjecture}\label{thm:general_0loss}
    Suppose there are $M$ players and $N$ arms.
    If all values of $S^\dagger_i$ are less than or equal to 1, it is possible to make the loss $L^\dagger$ equal to zero.
\end{conjecture}
This conjecture seems true, but so far, the proof has not been completed in generality and is left for future studies.

\section{Numerical Demonstrations}
In this section, we introduce several baseline models that output a joint probability selection matrix for given probabilistic preferences of two players.
Then, we show how much the loss can be improved by using the construction method of the optimal joint probability selection matrix introduced in Theorems \ref{thm:0loss_case} and \ref{thm:non0loss_case} (henceforth referred to as ``the optimal satisfaction matrix'').
The definitions of the notations, such as the preference and the loss, follow section \ref{subsec:problem}.
\subsection{Baselines}
\subsubsection{Uniform random}
In what we call ``uniform random'' method, the resulting joint selection probability matrix is such that all elements are equal except for the diagonals, which are filled with zeros.
That is to say, decision conflicts never happens, but the selection is determined completely randomly by the two players.
If we consider the preference setting shown in Table \ref{tb:pref_example}, the output of this method is
\begin{equation}
    \begin{pmatrix}
        0&\frac{1}{6}&\frac{1}{6}\\
        \frac{1}{6}&0&\frac{1}{6}\\
        \frac{1}{6}&\frac{1}{6}&0
    \end{pmatrix}.
\end{equation}
\begin{table}[t]
    \centering
    \caption{An Example of a preference setting}\label{tb:pref_example}
    \begin{tabular}{cccc}
        \hline
        Player&Arm 1&Arm 2&Arm 3\\
        \hline
        A&0.3&0.25&0.45\\
        B&0.5&0.2&0.3\\
        \hline
    \end{tabular}
\end{table}
\subsubsection{Simultaneous renormalization}
In this method, the product of each player's preference is considered first.
Then, the diagonals, where decision conflicts happen, are modified to zero, and finally, the whole joint selection probability matrix is renormalized so that the sum is 1.
The formula is as follows:
\begin{equation}
    \begin{gathered}
        r_{i,j} =
            \left\{
                \begin{array}{cc}
                A_i \cdot B_j & (i\neq j) \\
                0 & (i=j)
                \end{array}
            \right. ,\\
        p_{i,j} = \frac{r_{i,j}}{\sum\limits_{i,j} r_{i,j}}.
    \end{gathered}
\end{equation}
The joint selection probability matrix generated by this simultaneous renormalization method for the case in Table \ref{tb:pref_example} is
\begin{equation}
    \begin{pmatrix}
        0&0.06&0.09\\
        0.125&0&0.075\\
        0.225&0.09&0
    \end{pmatrix}/0.665 =
    \begin{pmatrix}
        0&0.090&0.1353\\
        0.1880&0&0.1128\\
        0.3383&0.1353&0
    \end{pmatrix}.
\end{equation}
\subsubsection{Random order}
In what we call ``random order'' method, the players first randomly determine in which order they will draw the arms.
This is inspired by random priority mechanism proposed by Abdulkadiro{\u{g}}lu {\it{et al.}} in the literature where preferences are deterministic \cite{abdulkadirouglu1998random}. Here instead, we consider probabilistic preference profiles.
Each player selects an arm according to the pre-determined order, but the arms already drawn by the previous players cannot be selected again; thus, the selection probabilities for those arms are set to zero.
Under the setting in Table \ref{tb:pref_example}, if we want to calculate $p_{1,2}$, two possible orders are considered. The first case is where player A draws first. In this case, the joint selection probability is
\begin{equation}
    0.3 \cdot \frac{0.2}{1-0.5} = 0.12.
\end{equation}
The other case is where player B draws first, and the probability is
\begin{equation}
    0.2 \cdot \frac{0.3}{1-0.25} = 0.08.
\end{equation}
Therefore, by taking the average,
\begin{equation}
    p_{1,2} = \frac{0.12 + 0.08}{2} = 0.1.
\end{equation}
Similarly, we can calculate all the joint selection probabilities and the resulting matrix is
\begin{equation}
    \begin{pmatrix}
        0&0.1&0.1718\\
        0.1674&0&0.1151\\
        0.3214&0.1243&0
    \end{pmatrix}.
\end{equation}
\subsection{Performance comparison}
\begin{figure}[t]
    \centering
    \includegraphics[width=13.0cm]{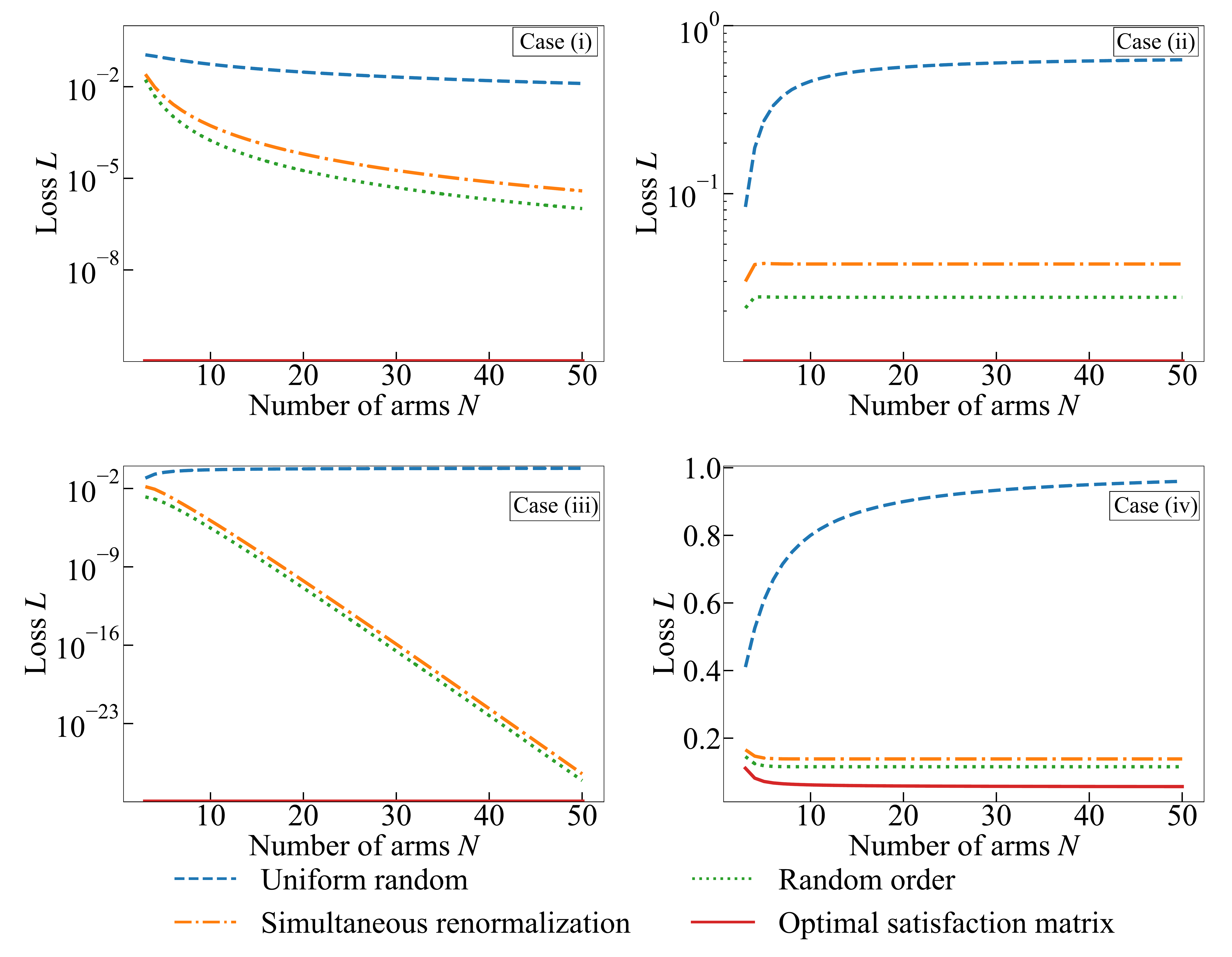}
    \caption{Loss comparison. $Y$ axes are log scale for case (i), (ii) and (iii). Lines for the optimal satisfaction matrix overlap with $X$ axes for these cases to show the loss is zero.
    (i)~arithmetic progression + same preference, (ii)~geometric progression with common ratio 2 + same preference, (iii)~geometric progression with common ratio 2 + reversed preference,
    (iv)~geometric progression with common ratio 3 + same preference. For case (i)--(iii), the minimum loss is zero, while for case (iv), it is greater than zero.}
    \label{fig:loss_comparison}
\end{figure}

We compare the loss $L$ of uniform random, simultaneous renormalization, random order and the optimal satisfaction matrix. Four preference settings shown below are examined to evaluate the performance of each method.
Namely, the degree of satisfaction of the players' preferences are investigated through the comparison of the loss $L$.
$c_i$ is used as a normalization term to ensure that the sum of the preference is 1.
\begin{enumerate}
    \item Arithmetic progression and same preference.\\
    $\displaystyle A_1: A_2:\cdots : A_N = B_1:B_2:\cdots :B_N = (1:2:\cdots : N) / c_1, \quad c_1 = \frac{(N+1)N}{2}$.
    \item Modified geometric progression with common ratio 2 and same preference.\\
    $\displaystyle A_1: A_2:\cdots : A_N = B_1:B_2:\cdots :B_N = (1:1:2:\cdots : 2^{N-2}) / c_2, \quad c_2 = 2^{N-1}$.
    \item Modified geometric progression with common ratio 2 and reversed preference.\\
    $\displaystyle A_1: A_2:\cdots : A_N = B_N:B_{N-1}:\cdots :B_1 = (1:1:2:\cdots : 2^{N-2}) / c_3, \quad c_3 = 2^{N-1}$.
    \item Geometric progression with common ratio 3 and same preference.\\
    $\displaystyle A_1: A_2:\cdots : A_N = B_1:B_2:\cdots :B_N = (1:3:\cdots : 3^{N-1}) / c_4, \quad c_4 = \frac{3^N-1}{2}$.
\end{enumerate}
Note that in case (i)--(iii), the optimal satisfaction matrix achieves $L=0$ since $\forall i;S_i \leq 1$, whereas in case (iv), it is not possible to achieve $L=0$ since
\begin{equation}
    S_N = 2\cdot \frac{3^{N-1}}{\sum_{i=0}^{N-1}3^i} = \frac{4\cdot 3^{N-1}}{3^N-1}>\frac{3^N}{3^N-1}>1.
\end{equation}
As for $N$, the following numbers are used.
\begin{equation}
    N = 3, 4, 5, \ldots, 50.
\end{equation}
Figure \ref{fig:loss_comparison} summarizes the loss $L$ as a function of the number of arms $N$ accomplished by each method.

In all cases, the optimal satisfaction matrix performs the best, followed by random order, simultaneous renormalization and uniform random.
The result in case (i) shows that the loss decreases as the number of arms rises for all the methods. This trend is due to our choice of the loss being similar to an $L_2$-norm. The absolute value of each preference $A_i, B_i$ becomes minor as the number of arms increases.

In the real world, the ratio of preference settings in case (ii) is likely to appear more frequently than in case (i). For uniform random, simultaneous renormalization, and random order, when two players have the same preference, the loss increases with the number of arms. In contrast, the result shows that the optimal satisfaction matrix consistently achieves 0-loss, which underlines the importance of the construction method of the optimal joint selection probability matrix in the real world, where there is a vast number of choices.

The result in case (iii) shows that, together with the optimal satisfaction matrix,  simultaneous renormalization and random order also have quite good accuracy. This result is due to the fact that the probability of decision conflicts happening is significantly smaller in this case (iii), where the players have strong, reversed preferences. The diagonal terms are renormalized in simultaneous renormalization, which gives perturbations to the other terms in the joint selection probability matrix. In this case (iii), these diagonal terms are smaller than in other cases, so the perturbations to the other terms are reduced. In random order, the second player sets his/her preference of the arm drawn by the first player to zero, but in the setup of case (iii), this preference tends to be small because the first player is more likely to select an arm with a higher preference, which is an unfavoured arm for the second player. This means that the second player can select an arm based on a preference that is almost equal to his/her original preference.

In case (iv), $S_N$ is greater than 1, so the optimal satisfaction matrix also does not achieve 0-loss. However, around $N=50$, the loss for random order is about 1.2 times smaller than the loss for simultaneous renormalization, while the optimal loss is almost twice smaller than the loss for random order.

\section{Conclusion}
In this study, we theoretically examined how to maximize each player's satisfaction by properly designing joint selection probabilities while avoiding decision conflicts when there are multiple players with probabilistic preferences.
In other words, the present study demonstrated how to accomplish conflict-free stochastic decision-making among multiple players wherein each player's preference is highly appreciated.
Particularly, we clarified the condition when the optimal joint selection probabilities perfectly eliminate the deviation of the resulting choice selection probabilities and the player's probabilistic preference in two-player, $N$-choice situations, which leads to what we call zero-loss realizations.
Furthermore, even under circumstances wherein zero-loss is unachievable, we showed how to construct, what we call, the optimal satisfaction matrix, whose joint selection probabilities minimize the loss. Moreover, we generalized the theory to $M$ player situations ($M \geq 3$), and proved the conditions wherein a zero-loss joint selection is impossible.
In addition, we numerically demonstrated the impact of the optimal satisfaction matrix by comparing several approaches that are able to provide conflict-free joint decision-making.

There are still many interesting future topics, including the mathematical proof of the conjecture shown in this study, which discusses the condition of zero-loss conflict-free joint stochastic selections involving an arbitrary number of players.
Detailed analysis of situations when we replace the loss with other metrics such as the KL divergence also remains to be explored.
Furthermore, extending the discussions to external environments, not just players' satisfaction, will be important in view of practical applications. This study paves a way toward multi-agent conflict-free stochastic decision-making.

\section*{Acknowledgments}
This work was supported in part by the CREST project (JPMJCR17N2) funded by the Japan Science and Technology Agency and Grants-in-Aid for Scientific Research (JP20H00233) funded by the Japan Society for the Promotion of Science.

\newpage

\bibliographystyle{RS}  
\bibliography{reference}  

\newpage

\renewcommand{\theequation}{S.\arabic{equation}}
\section*{Appendix A: Construction procedures for Assumption 2.1}
In the induction step of section \ref{sbsec:theorem1} of the main text, we assumed the existence of $p_{i,j}$s which satisfy certain conditions and proved that the perfect satisfaction is achievable using these $p_{i,j}$s.
Here, we prove this assumption; namely, Assumption 2.1 below is indeed true.
\setcounter{section}{2}
\begin{assumption}\label{assumption:Kth_arm}
    There exist $\hat{p}_{K,1}, \hat{p}_{K,2}, \ldots, \hat{p}_{K,N+1}, \hat{p}_{1,K}, \hat{p}_{2,K}, \ldots, \hat{p}_{N+1,K}$ which satisfy all the following conditions \eqref{eq_a}--\eqref{ineq_ab}.\\

    The sum of the $K$th row is equal to $\hat{A}_K$.
    \begin{equation}\label{eq_a}\sum_{j=1}^{N+1} \hat{p}_{K,j} = \hat{A}_K.\end{equation}

    The sum of the $K$th column is equal to $\hat{B}_K$.
    \begin{equation}\label{eq_b}\sum_{i=1}^{N+1} \hat{p}_{i,K} = \hat{B}_K.\end{equation}

    The sum of the $j$th column without the $K$th row is non-negative.
    \begin{equation}\label{ineq_a}\sum_{i \neq K} \hat{p}_{i,j} \geq 0 \Leftrightarrow \hat{p}_{K,j} \leq \hat{B}_j \quad (j \neq K).\end{equation}

    The sum of the $i$th row without the $K$th column is non-negative.
    \begin{equation}\label{ineq_b}\sum_{j \neq K} \hat{p}_{i,j} \geq 0 \Leftrightarrow \hat{p}_{i,K} \leq \hat{A}_i \quad (i \neq K).\end{equation}

    In the gray-shaded area of the joint selection probability matrix below, all the remaining popularities are smaller than or equal to the remaining total preference. Note that $\hat{S}_K = \min\{\hat{S}_i\}$.
    \begin{equation}\label{ineq_ab}(\hat{A}_i-\hat{p}_{i,K})+(\hat{B}_i-\hat{p}_{K,i}) \leq T-\hat{S}_K \quad (i \neq K).\end{equation}
\end{assumption}
Now, we can systematically determine $\hat{p}_{i,j}$s in the $K$th row or the $K$th column of the joint selection probability matrix to show Assumption \ref{assumption:Kth_arm} is correct; that is, these $\hat{p}_{i,j}$s fulfill the conditions \eqref{eq_a}--\eqref{ineq_ab}.
We cannot arbitrarily fill in the values in the $K$th row or the $K$th column so that their sum is $\hat{A}_K$ and $\hat{B}_K$, respectively, but have to take the other arms' preferences into consideration.
\eqref{ineq_a}, \eqref{ineq_b} and \eqref{ineq_ab} give us boundaries for $\hat{p}_{K,j}$s and $\hat{p}_{i,K}$s to exist.
Not only must we ensure that $\hat{p}_{K,j}$ and $\hat{p}_{i,K}$ do not exceed these boundaries, but we must also ensure that sufficient probability is assigned to the most popular arm $V$, because otherwise the left-hand side of \eqref{ineq_ab} can sometimes be greater than the right-hand side for arm $V$.
We consider the following three cases depending on size relation of $\hat{A}_K, \hat{B}_V, \hat{B}_K, \hat{A}_V$, and for each case, it is possible to construct the optimal $\hat{p}_{i,j}s$ that satisfy all the conditions \eqref{eq_a}--\eqref{ineq_ab}.
Note that these three cases are collectively exhaustive.
\begin{enumerate}[label=\textbf{[Case \arabic*]}]
    \item $\hat{A}_K \leq \hat{B}_V$ and $\hat{B}_K \leq \hat{A}_V$
    \item $\hat{A}_K > \hat{B}_V$
    \item $\hat{B}_K > \hat{A}_V$
\end{enumerate}
As a reminder, arm $K$ is the arm with the lowest popularity and arm $V$ is the one with the highest popularity. All the arms except for arm $V$ satisfies the following condition.
\begin{equation}\label{already_ineq_ab}
    \hat{S}_i \leq T - \hat{S}_K.
\end{equation}

\noindent\textbf{[Case 1]} $\hat{A}_K \leq \hat{B}_V$ and $\hat{B}_K \leq \hat{A}_V$

The following $\hat{p}_{i,j}$s on the $K$th row or $K$th column evidently satisfy the conditions \eqref{eq_a} and \eqref{eq_b}:
\begin{equation}\label{solution_situation1}
    \hat{p}_{K,V} = \hat{A}_K, \quad \hat{p}_{V,K}=\hat{B}_K, \quad \hat{p}_{K,j} =0 \quad (j\neq V), \quad \hat{p}_{i,K}=0 \quad (i\neq V).
\end{equation}
In addition, \eqref{ineq_a} and \eqref{ineq_b} are fulfilled because $\hat{A}_K \leq \hat{B}_V$ and $\hat{B}_K \leq \hat{A}_V$ in this particular case.
Moreover, \eqref{ineq_ab} is satisfied since for $i=V$,
\begin{equation}
    \hat{p}_{K,V}+\hat{p}_{V,K}=\hat{A}_K+\hat{B}_K=\hat{S}_K \geq \hat{S}_K-(T-\hat{S}_V) = \hat{S}_V+\hat{S}_K-T,
\end{equation}
which leads to
\begin{equation}
    (\hat{A}_V-\hat{p}_{V,K}) + (\hat{B}_V-\hat{p}_{K,V}) \leq T-\hat{S}_K.
\end{equation}
For the other $i$s, \eqref{ineq_ab} follows because they satisfy \eqref{already_ineq_ab} and \eqref{already_ineq_ab} is equivalent to \eqref{ineq_ab} when $\hat{p}_{K,i}=\hat{p}_{i,K}=0$.

Hence, $\hat{p}_{i,j}$s described in \eqref{solution_situation1} fulfill all the conditions \eqref{eq_a}--\eqref{ineq_ab} in \textbf{[Case 1]}.

\begin{table}[t]
    \centering
    \caption{An Example of [Case 1]}\label{tb:case1_example}
    \begin{tabular}{|c||c|c|c|c||c|}
        \hline
        Player&Arm 1&Arm 2&Arm 3&Arm 4&Total preference $T$\\
        \hline
        Player A&0.1&0.2&0.3&0.4&1.0\\
        Player B&0.2&0.2&0.1&0.5&1.0\\
        \hline\hline
        Popularity $S$&0.3&0.4&0.4&0.9&2.0\\
        \hline
    \end{tabular}
\end{table}
Table \ref{tb:case1_example} illustrates an example of such cases where $\hat{A}_K \leq \hat{B}_V$ and $\hat{B}_K \leq \hat{A}_V$.
Here, the most popular arm $K$ is arm 1 and the least popular arm $V$ is arm 4. In this case, the first row and the first column of the joint selection probability matrix should be filled in as follows:
\begin{equation}
    \left(\begin{array}{cccc}
        0&0&0&0.1\\
        0&\fillb *&\fillb *&\fillb *\\
        0&\fillb *&\fillb *&\fillb *\\
        0.2&\fillb *&\fillb *&\fillb *
    \end{array}\right)
\end{equation}
Then, the remaining gray-shaded region should satisfy the preference setting described in Table \ref{tb:case1_remain}.
Note that each popularity is less than or equal to the total preference in the remaining part; that is, the assumption of Lemma 2.1 holds for this part.
\begin{table}[t]
    \centering
    \caption{Preference profile for the remaining part in [Case 1]}\label{tb:case1_remain}
    \begin{tabular}{|c||c|c|c||c|}
        \hline
        Player&Arm 2&Arm 3&Arm 4&Total preference $T$\\
        \hline
        Player A&0.2&0.3&0.2&0.7\\
        Player B&0.2&0.1&0.4&0.7\\
        \hline\hline
        Popularity $S$&0.4&0.4&0.6&1.4\\
        \hline
    \end{tabular}
\end{table}

\noindent\textbf{[Case 2]} $\hat{A}_K > \hat{B}_V$

In this case, from the fact that $\hat{S}_K \leq \hat{S}_V$, it follows
\begin{equation}\label{cond_bk}
    \hat{B}_K < \hat{A}_V.
\end{equation}
When $\hat{B}_V+\hat{B}_1<\hat{A}_K$, let $m$ be the arm index which satisfies the following inequality.
\begin{equation}
    \hat{B}_V+\underbrace{\hat{B}_1+\hat{B}_2+\cdots+\hat{B}_{m-1}}_{\text{does not contain } \hat{B}_V \text{ or } \hat{B}_K}< \hat{A}_K \leq \hat{B}_V+\underbrace{\hat{B}_1+\hat{B}_2+\cdots+\hat{B}_{m}}_{\text{does not contain } \hat{B}_V \text{ or } \hat{B}_K}.
\end{equation}
When $\hat{B}_V+\hat{B}_1 \geq \hat{A}_K$, we define $m=1$.
$m$ always exists because
\begin{equation}
    \hat{B}_V+\underbrace{\hat{B}_1+\hat{B}_2+\cdots+\hat{B}_{N+1}}_{\text{does not contain } \hat{B}_V \text{ or } \hat{B}_K} = T-\hat{B}_K \geq \hat{A}_K.
\end{equation}
Then,
\begin{equation}\label{solution_situation2}
    \begin{cases}
        [\text{When }m=1]\\
        \hat{p}_{K,V}=\hat{B}_V, \quad \hat{p}_{K,1}=\hat{A}_K-\hat{B}_V, \quad \hat{p}_{K,j}=0 \quad (j \notin \{1, V\}),\\
        \hat{p}_{V,K}=\hat{B}_K, \quad \hat{p}_{i,K}=0 \quad (i \neq V)\\
        [\text{When }m \geq 2]\\
        \hat{p}_{K,K}=0, \quad \hat{p}_{K,V} = \hat{B}_V, \quad \hat{p}_{K,j}=\hat{B}_j \quad (j < m \text{ and } j \notin \{K, V\}),\\
        \hat{p}_{K,m}=\hat{A}_K-(\hat{B}_V+\underbrace{\hat{B}_1+\hat{B}_2+\cdots+\hat{B}_{m-1}}_{\text{does not contain } \hat{B}_V \text{ or } \hat{B}_K}), \quad \hat{p}_{K,j}=0 \quad (m < j \text{ and } j \notin \{K, V\}),\\
        \hat{p}_{V,K}=\hat{B}_K, \quad \hat{p}_{i,K}=0 \quad (i \neq V)
    \end{cases}
\end{equation}
evidently satisfy the conditions \eqref{eq_a} and \eqref{eq_b}.
Moreover, $\hat{p}_{i,j}$s in \eqref{solution_situation2} fulfill \eqref{ineq_a} because for $j=V, 1, 2, \ldots, m-1$,
\begin{equation}
    \hat{p}_{K,j} = \hat{B}_j \leq \hat{B}_j,
\end{equation}
and for $j=m$, the definition of $m$ has the condition $\hat{A}_K \leq \hat{B}_V+\underbrace{\hat{B}_1+\hat{B}_2+\cdots+\hat{B}_{m}}_{\text{does not contain } \hat{B}_V \text{ or } \hat{B}_K}$, which implies
\begin{equation}
    \hat{A}_K-(\hat{B}_V+\underbrace{\hat{B}_1+\hat{B}_2+\cdots+\hat{B}_{m-1}}_{\text{does not contain } \hat{B}_V \text{ or } \hat{B}_K}) \leq \hat{B}_m.
\end{equation}
For $j=m+1, m+2, \ldots, N+1$,
\begin{equation}
    \hat{p}_{K,j} = 0 \leq \hat{B}_j.
\end{equation}
Furthermore, with $\hat{p}_{i,j}$s described in \eqref{solution_situation2}, the condition \eqref{ineq_b} is also satisfied since for $i=V$,
\eqref{cond_bk} verifies
\begin{equation}
    \hat{p}_{V,K} = \hat{B}_K \leq \hat{A}_K,
\end{equation}
and for the other $i$s,
\begin{equation}
    \hat{p}_{i, K} = 0 \leq \hat{A}_i.
\end{equation}
Finally, \eqref{ineq_ab} is satisfied because for $i=V$,
\begin{align}\label{ineq_ab_situation2}
    \begin{split}
        \hat{p}_{K,V}+\hat{p}_{V,K} = \hat{B}_V + \hat{B}_K & \geq \hat{B}_V + \hat{B}_K - (T-\hat{A}_V-\hat{A}_K)\\
        &= \hat{S}_V+\hat{S}_K-T.
    \end{split}
\end{align}
Note that $T-\hat{A}_V-\hat{A}_K$ is always non-negative because $T \geq \hat{A}_V+\hat{A}_K$. \eqref{ineq_ab_situation2} is equivalent to
\begin{equation}
    (\hat{A}_V-\hat{p}_{V,K})+(\hat{B}_V-\hat{p}_{K,V}) \leq T-\hat{S}_K.
\end{equation}
For the other $i$s, they follow \eqref{already_ineq_ab}, and with the prerequisites $\hat{p}_{i,K} \geq 0, \quad \hat{p}_{K,i} \geq 0$, it follows \eqref{ineq_ab}.

Therefore, $\hat{p}_{i,j}$s given in \eqref{solution_situation2} satisfy all the conditions \eqref{eq_a}--\eqref{ineq_ab} in \textbf{[Case 2]}.

\begin{table}[t]
    \centering
    \caption{An Example of [Case 2]}\label{tb:case2_example}
    \begin{tabular}{|c||c|c|c|c||c|}
        \hline
        Player&Arm 1&Arm 2&Arm 3&Arm 4&Total preference $T$\\
        \hline
        Player A&0.25&0.1&0.15&0.5&1.0\\
        Player B&0.1&0.35&0.35&0.2&1.0\\
        \hline\hline
        Popularity $S$&0.35&0.45&0.5&0.7&2.0\\
        \hline
    \end{tabular}
\end{table}
Table \ref{tb:case2_example} shows an example of [Case 2].
Here, the most popular arm $K$ is arm 1 and the least popular arm $V$ is arm 4.
In this case, $m=2$ because
\begin{equation}
    \hat{B}_V(=0.2) < \hat{A}_K(=0.25) \leq \hat{B}_V(=0.2) + \hat{B}_2(=0.35).
\end{equation}
Therefore, the first row and the first column of the joint selection probability matrix should be filled in as follows:
\begin{equation}
    \left(\begin{array}{cccc}
        0&0.05&0&0.2\\
        0&\fillb *&\fillb *&\fillb *\\
        0&\fillb *&\fillb *&\fillb *\\
        0.1&\fillb *&\fillb *&\fillb *
    \end{array}\right)
\end{equation}
Then, the remaining gray-shaded region should satisfy the preference setting described in Table \ref{tb:case2_remain}.
Note that each popularity is less than or equal to the total preference in the remaining part; that is, the assumption of Lemma 2.1 holds for this part.
\begin{table}[t]
    \centering
    \caption{Preference profile for the remaining part in [Case 2]}\label{tb:case2_remain}
    \begin{tabular}{|c||c|c|c||c|}
        \hline
        Player&Arm 2&Arm 3&Arm 4&Total preference $T$\\
        \hline
        Player A&0.1&0.15&0.4&0.65\\
        Player B&0.3&0.35&0.0&0.65\\
        \hline\hline
        Popularity $S$&0.4&0.5&0.4&1.3\\
        \hline
    \end{tabular}
\end{table}

\noindent\textbf{[Case 3]} $\hat{B}_K > \hat{A}_V$

This case is quite similar to \textbf{[Case 2]}. If we swap $\hat{A}_i$ and $\hat{B}_i$ in the discussion in \textbf{[Case~2]},
we obtain $\hat{p}_{i,j}$s which satisfy all the conditions \eqref{eq_a}--\eqref{ineq_ab}.

\section*{Appendix B: Convexity of the loss $L$}
When we derived the optimal joint selection probability matrix in section \ref{sbsec:theorem2} of the main text, we used the fact that the loss function $L$ is convex.
Here, we prove that the loss function $L$ is indeed convex.
\begin{equation}
    L = \sum_i (\pi_A(i)-A_i)^2 + \sum_j (\pi_B(j)-B_j)^2
\end{equation}
We will prove that each $(\pi_A(i)-A_i)^2$ and $(\pi_B(j)-B_j)^2$ is convex, then we know that $L$ is convex because the sum of convex functions is also convex.

Now, the first term of $L$ is defined as
\begin{equation}
    L_1 := (\pi_A(1)-A_1)^2 = (p_{1,2}+p_{1,3}+\cdots+p_{1,N}-A_1)^2.
\end{equation}
The Hessian matrix for $L_1$ in terms of all $p_{i,j}$s $(i\neq j, i=1,2,\cdots,N, j=1,2,\cdots,N)$ is
\begin{equation}
    H_1 = \begin{pmatrix}D&O\\
    O&D_O\end{pmatrix}
\end{equation}
where
\begin{equation}
    D = \overbrace{\begin{pmatrix}2\quad 2\quad \cdots\quad 2\\
    2\quad 2\quad \cdots\quad 2\\
    \vdots\quad \vdots\quad \vdots\quad \vdots \\
    2\quad 2\quad \cdots\quad 2\end{pmatrix}}^{N-1},\quad D_O = \overbrace{\begin{pmatrix}0\quad 0\quad \cdots\quad 0\\
    0\quad 0\quad \cdots\quad 0\\
    \vdots\quad \vdots\quad \vdots\quad \vdots \\
    0\quad 0\quad \cdots\quad 0\end{pmatrix}}^{(N-1)^2}.
\end{equation}
Since this is a block diagonal matrix, the eigenvalues of $H_1$ is the union of the eigenvalues of $D$ and $D_o$.
The eigenvalues of $D_o$ are obviously $\overbrace{0,0,\cdots ,0}^{(N-1)^2}$ and those of $D$ are $2(N-1), \overbrace{0,0,\cdots,0}^{N-2}$.

Therefore, all the eigenvalues of $H_1$ are non-negative, which means that $H_1$ is positive semidefinite, and this implies $L_1$ is convex. Similarly, each $(\pi_A(i)-A_i)^2$ and $(\pi_B(j)-B_j)^2$ is proven to be convex because of the symmetry of the loss function, given that $i=1$ is not special among all $i$s.
From the above argument, we now know that the loss function $L$ is convex.

\section*{Appendix C: Verification of the KKT conditions}
In section \ref{sbsbsec:KKT} of the main text, where we derived the point which satisfies all the KKT conditions, we defined new notations and functions.
The flattened vector of a joint selection probability matrix is defined as
\begin{equation}
    \boldsymbol{p} = \begin{pmatrix}p_{1,2}&p_{1,3}&\cdots&p_{1,N}&p_{2,1}&p_{2,3}&\cdots&p_{2,N}&\cdots&p_{N,N-1}\end{pmatrix}^T
\end{equation}
Functions $h$ and $g_{i,j}$s are defined as follows:
\begin{equation}
    h(\boldsymbol{p}) = \sum_{i,j} p_{i,j}-1, \quad g_{i,j}(\boldsymbol{p}) = -p_{i,j}\quad (i\neq j).
\end{equation}
Now, the problem can be written as
\begin{equation}
    \begin{aligned}
    \min_{\boldsymbol{p}} \quad & L(\boldsymbol{{p}})\\
    \textrm{s.t.} \quad & h(\boldsymbol{p}) = 0, \quad g_{i,j}(\boldsymbol{p})\leq 0.
    \end{aligned}
\end{equation}
$L(\boldsymbol{p})$ is the loss which corresponds to $\boldsymbol{p}$. Since the objective function and the constraints are all convex, $\tilde{\boldsymbol{p}}$ which satisfies the KKT conditions below gives the global minimum.
\begin{gather}
    \nabla L(\tilde{\boldsymbol{p}}) + \sum_{i,j} \lambda_{i,j} \nabla g_{i,j}(\tilde{\boldsymbol{p}}) + \mu \nabla h(\tilde{\boldsymbol{p}}) = \boldsymbol{0},\label{stationarity}\\
    \lambda_{i,j} g_{i,j}(\tilde{\boldsymbol{p}}) = 0,\label{slackness}\\
    \lambda_{i,j} \geq 0,\label{dual}\\
    g_{i,j}(\tilde{\boldsymbol{p}}) \leq 0, \quad h(\tilde{\boldsymbol{p}})=0.\label{primal}
\end{gather}
Here, we will show that
\begin{gather}
    \epsilon = \frac{S_N-1}{2(N-1)},\label{opt_eps}\\
    \mu = 2(N-2)\epsilon ,\label{opt_mu}\\
    \lambda_{i,j} = \left\{
            \begin{array}{cl}
            2N\epsilon & (i\neq N \text{ and } j\neq N) \\
            0 & (otherwise)
            \end{array}
    \right.,\label{opt_lambda}\\
    \tilde{p}_{i,j} = \left\{
        \begin{array}{cl}
        0 & (i\neq N \text{ and } j\neq N) \\
        A_i + \epsilon & (j=N)\\
        B_j + \epsilon & (i=N)
        \end{array}
    \right.\label{opt_p}
\end{gather}
satisfy \eqref{stationarity}--\eqref{primal}.

\noindent\textbf{[Condition 1] Stationarity \eqref{stationarity}.}

In the following argument, we call the element in each vector which is in the same dimension as $p_{i,j}$ ``the $(i,j)$th element'' in a vector.

From the definition of $L$, the $(i,j)$th element of $\nabla L(\boldsymbol{p})$, denoted by $\nabla L(\boldsymbol{p})_{[i,j]}$, is
\begin{equation}
    \nabla L(\boldsymbol{p})_{[i,j]} = 2(\pi_A(i) - A_i) + 2(\pi_B(j) - B_j).
\end{equation}
With $\tilde{p}_{i,j}$s described in \eqref{opt_p},
\begin{equation}
    \left(\nabla L(\tilde{\boldsymbol{p}})\right)_{[i,j]} = \left\{
            \begin{array}{cl}
                2\epsilon + 2\epsilon = 4\epsilon & (i\neq N \text{ and } j\neq N) \\
                -2(N-1)\epsilon + 2\epsilon = -2(N-2)\epsilon & (otherwise)
            \end{array}
    \right. .
\end{equation}
Also, with $\tilde{p}_{i,j}$s and $\lambda_{i,j}$s described in \eqref{opt_lambda} and \eqref{opt_p}, the $(i,j)$th element of $\displaystyle \sum\limits_{i,j} \lambda_{i,j} \nabla g_{i,j}(\tilde{\boldsymbol{p}})$ is
\begin{equation}
    \left(\displaystyle \sum\limits_{i,j} \lambda_{i,j} \nabla g_{i,j}(\tilde{\boldsymbol{p}})\right)_{[i,j]} = \left\{
        \begin{array}{cl}
            -2N\epsilon & (i\neq N \text{ and } j\neq N) \\
            0 & (otherwise)
        \end{array}
\right. .
\end{equation}
Moreover, the $(i,j)$th element of $\mu \nabla h(\tilde{\boldsymbol{p}})$ is
\begin{equation}
    \left(\mu \nabla h(\tilde{\boldsymbol{p}})\right)_{[i,j]} = 2(N-2)\epsilon.
\end{equation}
Therefore, stationarity is successfully achieved with $\tilde{p}_{i,j}$s, $\lambda_{i,j}$s and $\mu$ given in \eqref{opt_eps}--\eqref{opt_p}.
\begin{equation}
    \nabla L(\tilde{\boldsymbol{p}}) + \sum_{i,j} \lambda_{i,j} \nabla g_{i,j}(\tilde{\boldsymbol{p}}) + \mu \nabla h(\tilde{\boldsymbol{p}}) = \boldsymbol{0}.
\end{equation}

\noindent\textbf{[Condition 2] Complementary slackness \eqref{slackness}.}

From \eqref{opt_lambda} and \eqref{opt_p}, in both cases, where $i\neq N \land j\neq N$ or not,
\begin{equation}
    \lambda_{i,j}g_{i,j}(\tilde{\boldsymbol{p}}) = -\lambda_{i,j}\tilde{p}_{i,j} = 0 \geq 0.
\end{equation}

\noindent\textbf{[Condition 3] Dual feasibility \eqref{dual}.}

Since $S_N-1 > 0$, it is evident that $\epsilon$ is positive and so is $2N\epsilon$. Thus,
\begin{equation}
    \lambda_{i,j} \geq 0.
\end{equation}

\noindent\textbf{[Condition 4] Primal feasibility \eqref{primal}.}

Since $\epsilon > 0$, it follows that $A_i+\epsilon$ and $B_j+\epsilon$ are non-negative. Then,
\begin{equation}
    g_{i,j}(\tilde{\boldsymbol{p}})\leq 0.
\end{equation}
Also,
\begin{align}
    \begin{split}
        h(\tilde{\boldsymbol{p}}) &= \sum_{i,j} p_{i,j}-1 \\
        &= \sum_{i=1}^{N-1}(A_i+\epsilon) + \sum_{j=1}^{N-1}(B_j+\epsilon)-1\\
        &= \sum_{i=1}^{N-1}(A_i+B_i) + 2(N-1)\epsilon -1\\
        &= 2-S_N + (S_N-1)-1 = 0.
    \end{split}
\end{align}
Therefore,
\begin{equation}
    h(\tilde{\boldsymbol{p}}) = 0
\end{equation}
holds.

\end{document}